\documentstyle[12pt]{article}

\newcommand{\EQ}{\begin{equation}}
\newcommand{\EN}{\end{equation}}
\newcommand{\bea}{\begin{eqnarray}}
\newcommand{\ena}{\end{eqnarray}}
\newcommand{\vs}[1]{\vspace{#1 mm}}

\renewcommand{\a}{\alpha}

\renewcommand{\d}{\delta}
\newcommand{\e}{\epsilon}
\def\bbox{{\,\lower0.9pt\vbox{\hrule \hbox{\vrule height 0.2 cm
\hskip 0.2 cm \vrule height 0.2 cm}\hrule}\,}}
\newcommand{\dsl}{\pa \kern-0.5em /}
\newcommand{\la}{\lambda}
\newcommand{\shalf}{\frac{1}{2}}
\newcommand{\pa}{\partial}
\renewcommand{\t}{\theta}

\newcommand{\nn}{\nonumber\\}

\begin{document}
\topmargin 0pt
\oddsidemargin 0mm
\renewcommand{\thefootnote}{\fnsymbol{footnote}}
\begin{titlepage}
\setcounter{page}{0}
\begin{flushright}
OU-HET 277 \\
hep-th/9709065
\end{flushright}

\vs{10}
\begin{center}
{\Large
Realization of D4-Branes at Angles in Super Yang-Mills Theory
}
\vs{10}

{\large
Nobuyoshi Ohta\footnote{e-mail address: ohta@phys.wani.osaka-u.ac.jp}
and Jian-Ge Zhou\footnote{e-mail address:
jgzhou@phys.wani.osaka-u.ac.jp, JSPS postdoctral fellow}} \\
\vs{5}
{\em Department of Physics, Osaka University, \\
Toyonaka, Osaka 560, Japan}
\end{center}
\vs{5}

\centerline{{\bf{Abstract}}} \vs{5}
An explicit solution in super Yang-Mills theory, which after T-duality
describes two sets of D4-branes at angles, is constructed. The gauge
configuration possesses 3/16 unbroken supersymmetry for the equal
magnitudes of field strengths, and can be considered as the counterpart of
the solution of $D=11$ supergravity with the same amount of supersymmetry
in the solutions given by Gauntlett et al. The energy of the Born-Infeld
action of the gauge configuration gives further evidence for the
geometrical interpretation as two sets of D4-branes at angles. The energy
of super Yang-Mills theory is shown to coincide with that of M(atrix)
theory. This fact shows that the configuration with 3/16 supersymmetry
can be realized in M(atrix) theory, which describes two sets of
longitudinal M5-branes (with common string direction) at angles.
\vs{5}



\end{titlepage}
\newpage
\renewcommand{\thefootnote}{\arabic{footnote}}
\setcounter{footnote}{0} 

Recently D-branes~\cite{POL} have played a crucial role in revealing
the nonperturbative structure of string theories and M theory.
The low-energy effective dynamics of $N$ coincident D-branes may be
described by super $U(N)$ Yang-Mills theory~\cite{W1}. Furthermore,
an interesting proposal has been put forward that a microscopic
description of the M theory is provided by the large $N$ limit of
the compactified super Yang-Mills in the infinite momentum frame~\cite{BFS}.
Thus a profound connection is emerging between string theory and gauge
theories.

In fact, with the identification of D-branes with various RR solitons, the
properties and degeneracies of BPS states can be reflected in super
Yang-Mills theory. The connection between D-branes and torons (instantons
on a torus) was studied in~\cite{GR1}-\cite{GR}, where the gauge
configurations have 1/2 or 1/4 unbroken supersymmetry.
T-duality turns those to gauge configurations carrying only D2-brane
charges, which relates the constraints on the existence of torons to
geometrical constraints on the supersymmetric configurations of intersecting
D2-branes at angels~\cite{GR1}-\cite{BL}. Other related works on branes
at angles preserving half or quarter supersymmetry have been given
either from the string theory point of view in~\cite{LI} or from supergravity
in~\cite{BM}-\cite{HA}. However, how to construct a super Yang-Mills solution
on the torus $T^8$, which after T-duality corresponds to intersecting
D4-branes at angels, is unclear.

On the other hand, a configuration of non-orthogonal two M5-branes
with common string direction, as the solution of $D=11$ supergravity,
was discussed in~\cite{GG,TO}. A remarkable feature of this solution is
that it preserves 3/16 unbroken supersymmetry, a fraction not obtainable
from orthogonal intersections~\cite{PT}-\cite{O}. The 3/16 supersymmetry
derives from the $Sp(2)$ holonomy of hyper-K\"{a}hler 8-metrics. In the
case of the IIA D4-branes, a pair of intersecting D4-branes preserves 3/16
supersymmetry if their orientations are related by a rotation in an
$Sp(2)$ subgroup of $SO(8)$ commuting with multiplication by a quaternion.
Specifically, the corresponding rotation matrix in the spinor
representation can be chosen as~\cite{TO}
\EQ
R= \exp\left[ \shalf\theta \left( \Gamma_{12} + \Gamma_{34} +
\Gamma_{56} +\Gamma_{78} \right) \right],
\label{rot}
\EN
where it has been assumed that the D4-branes originally lie on 1357
directions, and the first D4-brane is rotated away from
1357 D4-brane along 2468 directions.

Now a natural question appears whether the above solution of $D=11$
supergravity with 3/16 supersymmetry has a counterpart in super
Yang-Mills theory on $T^8$. In order to answer this question, we propose
a special constant background field configuration in super Yang-Mills
theory on $T^8$, which, as we will see, breaks gauge group $U(N)$ to
$U(l)\times U(k)$, $(l+k=N)$, and the resulting bound states may be
denoted as \{8666644\}. The meaning of this notation is that it has one
8-brane charge corresponding to the 8-brane wrapped around the compact
spatial directions (12345678), four 6-brane charges corresponding to
6-branes wrapped on (345678), (125678), (123478) and (123456) directions,
respectively, and two 4-brane charges corresponding to the 4-branes
wrapped along (5678) and (1234) directions. Given this \{8666644\}
configuration, we can use T-duality to relate it to a configuration
carrying only 4-brane charges, in which one set of D4-branes is rotated
off the (1357) 4-plane by rotations in the (12) and (34) planes, and the
other set is rotated away from the (1357) 4-plane by the rotations in the
(56) and (78) planes. By taking into account both the linear and nonlinear
supersymmetries of the D-brane worldvolume theory, we find that
the gauge configuration with the equal field strengths
$B_{12}=B_{34}=-B_{56}=-B_{78}$ preserves
3/16 unbroken supersymmetry indeed, but the other with $B_{12}=B_{34}$
and $B_{56}=B_{78}$ preserves only 1/8 supersymmetry.

The energy of the gauge configuration is calculated by using the
Born-Infeld action, which agrees with that obtained from geometrical
consideration. The agreement shows that the bound state is a generalized
``non-marginal" one, whose mass has the form
\EQ
m \sim \sqrt{(q_1^{(1)})^2 + (q_1^{(2)})^2 + (q_1^{(3)})^2 + (q_1^{(4)})^2}
 + \sqrt{(q_2^{(1)})^2 + (q_2^{(2)})^2 + (q_2^{(3)})^2 + (q_2^{(4)})^2},
\EN
where $q$'s are brane charges. This gives further evidence that
after T-duality the gauge configuration
can be interpreted as two sets of D4-branes at angles. To analyse the
present configuration in the framework of the M(atrix) theory, the
energy of super Yang-Mills theory is evaluated. It is found that the energy
of super Yang-Mills theory matches that obtained from the M(atrix) theory
in the $N \to \infty$ limit. This observation indicates that the solution
with 3/16 unbroken supersymmetry can be realized in M(atrix) theory, which
describes two sets of longitudinal M5-branes at relative angles with
common string direction.

Let us start by considering the configuration with the backgrounds
\bea
A_1=0, && A_2=F_{12}x_1, \nn
A_3=0, && A_4=F_{34}x_3, \nn
A_5=0, && A_6=F_{56}x_5, \nn
A_7=0, && A_8=F_{78}x_7,
\ena
on $T^8$ with sides of lengths $a_i, i=1,\cdots,8$. For the field strengths,
we take constant diagonal ones which break $U(N)$ to $U(l)\times U(k)$,
($l+k=N$), where the nonzero $U(N)$ fields are of the form
\bea
F_{12}=B_{12}\: {\cal U}_l, && F_{34}=B_{34}\: {\cal U}_l, \nn
F_{56}=B_{56}\: {\cal U}_k, && F_{67}=B_{78}\: {\cal U}_k,
\label{fi}
\ena
with
\bea
{\cal U}_l &=& {\rm diag.} \left(1_{l\times l} ,0_{k\times k} \right), \nn
{\cal U}_k &=& {\rm diag.} \left(0_{l\times l} ,1_{k\times k} \right).
\label{mat}
\ena
By considering translations in 1, 3, 5, 7 directions, we can derive
the flux quantization conditions~\cite{GR1}
\bea
B_{12}a_1a_2 = 2 \pi \frac{n_{12}^{(l)}}{l}, &&
B_{34}a_3a_4 = 2 \pi \frac{n_{34}^{(l)}}{l}, \nn
B_{56}a_5a_6 = 2 \pi \frac{n_{56}^{(k)}}{k}, &&
B_{78}a_7a_8 = 2 \pi \frac{n_{78}^{(k)}}{k},
\label{qu}
\ena
where $n_{12}^{(l)}, n_{34}^{(l)}, n_{56}^{(k)}, n_{78}^{(k)}$ are
nonzero twists defined in~\cite{TH}.

{}From the Chern-Simons couplings of field strengths to the RR
potentials~\cite{W2,DO}, we can show that the above configuration describes
the bound state of \{8666644\}. It consists of one 8-brane (12345678); four
6-branes (345678), (125678), (123478) and (123456); and two 4-branes
(5678) and (1234); and the corresponding charges are $N, n_{12}^{(l)},
n_{34}^{(l)}, n_{56}^{(k)}, n_{78}^{(k)}, n_{12}^{(l)} n_{34}^{(l)}/l$
and $n_{56}^{(k)} n_{78}^{(k)}/k$, respectively.

Describing $A_\mu$ as a connection on a bundle with trivial boundary
conditions in directions 2, 4, 6 and 8, we can T-dualize in those
directions to arrive at a configuration comprising two sets of
D4-branes. In each set, there are four kinds of D4-branes.
The embeddings of these D4-branes are given by~\cite{POL,TA1}
\bea
X_2 = \frac{F_{12}x_1}{2 \pi}, &&
X_4 = \frac{F_{34}x_3}{2 \pi}, \nn
X_6 = \frac{F_{56}x_5}{2 \pi}, &&
X_8 = \frac{F_{78}x_7}{2 \pi},
\label{4b}
\ena
where $(2\pi)^2 \a'=1$ is chosen. We can see that the T-dualized configuration
describes the bound state of D4-branes denoted by (1357), (2357),
(1457), (1367), (1358), (2457) and (1368), which are wrapped on the dual
${\hat T}^8$ with the lengths $b_2=1/a_2, b_4=1/a_4, b_6=1/a_6, b_8=1/a_8$
for directions 2, 4, 6 and 8, and $b_1=a_1, b_3=a_3, b_5=a_5, b_7=a_7$
for 1, 3, 5 and 7 directions. In fact, these D4-branes given in
eq.~(\ref{4b}) can be divided into two sets described by
\bea
X_2^{(l)} = \frac{B_{12} x_1}{2\pi}, \;\;
X_4^{(l)} = \frac{B_{34} x_3}{2\pi}, \;\;
X_6^{(l)} = 0, \;\;
X_8^{(l)} = 0, \nn
X_2^{(k)} = 0, \;\;
X_4^{(k)} = 0, \;\;
X_6^{(k)} = \frac{B_{56} x_5}{2\pi}, \;\;
X_8^{(k)} = \frac{B_{78} x_7}{2\pi}.
\label{fs}
\ena
In each set, there are four sorts of D4-branes:

Set A: (1357), (2357), (1457), (2457),

Set B: (1357), (1367), (1358), (1368), \\
and the corresponding charges are given by
\bea
Q_{1357}^{(l)} = l, \;\;
Q_{2357}^{(l)} = n_{12}^{(l)}, \;\;
Q_{1457}^{(l)} = n_{34}^{(l)}, \;\;
Q_{2457}^{(l)} = \frac{n_{12}^{(l)}n_{34}^{(l)}}{l}, \nn
Q_{1357}^{(k)} = k, \;\;
Q_{1367}^{(k)} = n_{56}^{(k)}, \;\;
Q_{1358}^{(k)} = n_{78}^{(k)}, \;\;
Q_{1368}^{(k)} = \frac{n_{56}^{(k)}n_{78}^{(k)}}{k}.
\label{nu}
\ena

These D4-branes can be interpreted as a system obtained by two sets
of D4-branes, the projections of whose charges onto the (1357) plane are
$l$ and $k$, respectively. One of them is rotated off the (1357) plane
by rotations in the (12) and (34) planes. The other set is rotated away
from the (1357) plane by the rotations in the (56) and (78) planes. The
angles $\t_{12}, \t_{34}, \t_{56}$ and $\t_{78}$, which mix the directions
(12), (34), (56) and (78), respectively, can be defined geometrically as
\bea
\tan\t_{12} = \frac{X_2^{(l)}}{x_1} = \frac{B_{12}}{2\pi}, \;
&& \tan\t_{34} = \frac{X_4^{(l)}}{x_3} = \frac{B_{34}}{2\pi}, \nn
\tan\t_{56} = \frac{X_6^{(k)}}{x_5} = \frac{B_{56}}{2\pi}, \;
&& \tan\t_{78} = \frac{X_8^{(k)}}{x_7} = \frac{B_{78}}{2\pi},
\label{an1}
\ena
where $0<B_{2i-1,2i}<\infty, i=1,\cdots,4$, {\it i.e.},
$0<\t_{2i-1,2i}<\frac{\pi}{2}$.

The volumes for D4-branes can be calculated from eqs.~(\ref{fs}) and
(\ref{nu}) with
\bea
\label{11}
V_{1357}^{(l)} = l b_1 b_3 b_5 b_7, \; &&
V_{2357}^{(l)} = n_{12}^{(l)} b_2 b_3 b_5 b_7, \nn
V_{1457}^{(l)} = n_{34}^{(l)} b_1 b_4 b_5 b_7, \; &&
V_{2457}^{(l)} = \frac{n_{12}^{(l)}n_{34}^{(l)}}{l} b_2 b_4 b_5 b_7, \\
V_{1357}^{(k)} = k b_1 b_3 b_5 b_7, \; &&
V_{1367}^{(k)} = n_{56}^{(k)} b_1 b_3 b_6 b_7, \nn
V_{1358}^{(k)} = n_{78}^{(k)} b_1 b_3 b_5 b_8, \; &&
V_{1368}^{(k)} = \frac{n_{56}^{(k)}n_{78}^{(k)}}{k} b_1 b_3 b_6 b_8.
\label{12}
\ena
Eq.~(\ref{11}) can be understood as the projections of the volume of
the first set of rotated D4-branes along (1357), (2357), (1457)
and (2457), and eq.~(\ref{12}) has a similar explanation.

The above interpretation gives another definition for the angles
$\t_{12}, \t_{34},\t_{56}$ and $\t_{78}$:
\bea
\tan\t_{12} = \frac{V_{2357}^{(l)}}{V_{1357}^{(l)}}, \;
&& \tan\t_{34} = \frac{V_{1457}^{(l)}}{V_{1357}^{(l)}}, \nn
\tan\t_{56} = \frac{V_{1367}^{(k)}}{V_{1357}^{(k)}}, \;
&& \tan\t_{78} = \frac{V_{1358}^{(k)}}{V_{1357}^{(k)}}.
\label{an2}
\ena
The combination of eqs.~(\ref{an1}) and (\ref{an2}) gives nothing but
the flux quantization conditions in eq.~(\ref{qu}).

If the geometrical interpretation as two sets of D4-branes at angles
is correct, the total volume of the system can be written as
\EQ
V = \sqrt{ \left[V_{1357}^{(l)}\right]^2 + \left[V_{2357}^{(l)}\right]^2
 + \left[V_{1457}^{(l)}\right]^2 + \left[V_{2457}^{(l)}\right]^2}
 + \sqrt{ \left[V_{1357}^{(k)}\right]^2 + \left[V_{1367}^{(k)}\right]^2
 + \left[V_{1358}^{(k)}\right]^2 + \left[V_{1368}^{(k)}\right]^2}.
\EN
By eqs.~(\ref{qu}), (\ref{11}) and (\ref{12}), this can be cast into
\EQ
V = a_1 a_3 a_5 a_7 \left\{ l \sqrt{
 \left[1+\left(\frac{B_{12}}{2\pi}\right)^2 \right]
 \left[1+\left(\frac{B_{34}}{2\pi}\right)^2 \right] }
 + k \sqrt{
 \left[1+\left(\frac{B_{56}}{2\pi}\right)^2 \right]
 \left[1+\left(\frac{B_{78}}{2\pi}\right)^2 \right] } \right\}.
\EN
The mass of the bound state of D4-branes is then given by multiplying with
4-brane tension $T_4$:
\bea
m &=& T_4 V \nn
&=& \frac{2\pi}{g_4} a_1 a_3 a_5 a_7 \left\{ l \sqrt{
 \left[1+\left(\frac{B_{12}}{2\pi}\right)^2 \right]
 \left[1+\left(\frac{B_{34}}{2\pi}\right)^2 \right] }
 + k \sqrt{
 \left[1+\left(\frac{B_{56}}{2\pi}\right)^2 \right]
 \left[1+\left(\frac{B_{78}}{2\pi}\right)^2 \right] } \right\},\nn
\label{mass}
\ena
where we have chosen $\a'=(2\pi)^{-2}$, and $g_4$ is the string coupling
on the dual torus ${\hat T}^8$.

The above volume expression for two sets of D4-branes with relative angles
can be shown correct by considering the energy of the Born-Infeld
action for \{8666644\} configuration~\cite{HT,TA1}
\EQ
E_{BI} = T_8 \;{\rm Tr} \int d^8 x \sqrt{-\det \left(
 \eta_{\mu\nu} + 2\pi \a' F_{\mu\nu} \right) }.
\EN
From eqs.~(\ref{fi}) and (\ref{mat}), we find that this expression gives
\EQ
E_{BI} = T_8 \prod_{i=1}^8 a_i \left\{ l \sqrt{
 \left[1+\left(\frac{B_{12}}{2\pi}\right)^2 \right]
 \left[1+\left(\frac{B_{34}}{2\pi}\right)^2 \right] }
 + k \sqrt{
 \left[1+\left(\frac{B_{56}}{2\pi}\right)^2 \right]
 \left[1+\left(\frac{B_{78}}{2\pi}\right)^2 \right] } \right\},
\EN
with
\EQ
T_8 = \frac{2\pi}{g_8}, \;\;
g_4 = \frac{g_8}{a_2 a_4 a_6 a_8},
\EN
where $g_8$ is the string coupling on $T^8$.

Comparing $m$ with $E_{BI}$, we find that they are equal, which gives
further evidence for the geometrical interpretation of two sets of D4-branes
with relative angles.

Before analysing the unbroken supersymmetry, one might be tempted to
consider some kind of self-duality conditions in eight dimensions of
the type
\EQ
F_{\mu\nu} = T_{\mu\nu\rho\sigma} F^{\rho\sigma},
\EN
with $T_{\mu\nu\rho\sigma}$ a certain fixed four-form in eight dimensions.
A large literature is devoted to this subject (see e.g.~\cite{BKS,HO}, and
references therein). Self-duality in eight dimensions indeed exhibits
some remarkable properties: group-theoretical classification of
possible four-forms $T_{\mu\nu\rho\sigma}$ has been presented, and an
analogy of the ADHM construction has been shown to exist in some cases;
the algebra of octonions plays a prominent role in some of these
constructions. In the present case, we interpret the T-dualized
configuration as rotations of two sets of D4-branes by an $Sp(2)$ subgroup
of $SO(8)$, which was shown to be related via M-theory dualities to
8-dimensional hyper-K\"{a}hler manifolds, and this can be realized by
choosing~\cite{GR1,BKS}
\EQ
F_{12}^{SU(N)} = F_{34}^{SU(N)} = F_{56}^{SU(N)} = F_{78}^{SU(N)},
\label{su}
\EN
where
\EQ
F_{ij}^{SU(N)} = F_{ij}^{U(N)} - \frac{1}{N} {\rm Tr} F_{ij}^{U(N)}.
\label{u}
\EN
Combining eqs.~(\ref{su}) and (\ref{u}), one gets
\EQ
B_{12} = B_{34} = - B_{56} = - B_{78},
\label{eq}
\EN
which shows that the first set of D4-branes is rotated off (1357)
along (12) and (34) planes with an angle $\t$, and the second set is
rotated away from (1357) along (56) and (78) planes with the angle $-\t$.

By taking into account both the linear and nonlinear supersymmetries
of the D-brane worldvolume theory~\cite{BSS}, we have
\EQ
\d \la = \Gamma^{\mu\nu} F _{\mu\nu} \e + {\tilde \e} =0.
\EN
From the expression for $F_{\mu\nu}$, we are lead to
\bea
(B_{12} \Gamma^{12} + B_{34} \Gamma^{34}) \e + {\tilde \e} = 0, \nn
(B_{56} \Gamma^{56} + B_{78} \Gamma^{78}) \e + {\tilde \e} = 0,
\label{cond1}
\ena
which can be transformed by eq.~(\ref{eq}) into
\bea
{\tilde \e} &=& - B_{12} (\Gamma^{12} + \Gamma^{34}) \e, \nn
\Sigma \e &=& 0,
\ena
with
\EQ
\Sigma = \Gamma^{12} + \Gamma^{34} + \Gamma^{56} + \Gamma^{78}.
\EN

As discussed in~\cite{TO}, the eigenvalues of $\Sigma^2$ are
\EQ
-4 (8,-), \;\;
-16(2,+), \;\;
0(6,+).
\EN
The numbers in the parentheses are the multiplicities of the eigenvalues
and the signs are those of the eigenvalues of $\Gamma_{12345678}$.
Of course, the zero eigenvalues of $\Sigma^2$ correspond to zero
eigenvalues of $\Sigma$. Therefore there exist precisely 6 nonzero
components of $\e$, which implies preservation of
$\frac{6}{16}\times \shalf = \frac{3}{16}$ supersymmetry.

When $B_{12} = B_{34}, B_{56} = B_{78}$, but $B_{12}$ and $B_{56}$ are
two independent variables, the condition (\ref{cond1}) turns into
\EQ
(\Gamma^{12} + \Gamma^{34}) \e=0, \;\;
(\Gamma^{56} + \Gamma^{78}) \e=0, \;\;
{\tilde \e} =0,
\label{cond2}
\EN
which shows that the configuration preserves 1/8 supersymmetry.
This corresponds to the case that the rotation matrix in the spinor
representation has the form 
\EQ
R= \exp\left[ \shalf\theta \left( \Gamma_{12} + \Gamma_{34}\right) +
 \shalf\psi \left( \Gamma_{56} +\Gamma_{78} \right) \right],
\EN
where $\t$ and $\psi$ are generic angles.

{}From the above discussion, we know that the first set of D4-branes rotates
along (12) and (34) with angle $\t$ and the second set along (56) and (78)
with angle $-\t$, which is slightly different from that in~\cite{GG,TO}
where the first D4-brane rotates along (12), (34), (56) and (78) planes
with the same angle $\t$, but the second does not rotate. In fact, the
spinor representation of the rotation matrix for the present
configuration in eq.~(\ref{eq}) has the form~\cite{BL}
\EQ
R= \exp\left[ \shalf{\cal U}_l \theta \left( \Gamma_{12} + \Gamma_{34}\right)
 - \shalf{\cal U}_k \theta \left( \Gamma_{56} +\Gamma_{78} \right) \right],
\label{rot2}
\EN
which is different from that in~\cite{TO}. However, after redefining the
killing spinor, we find that the constraints are equivalent. Rewriting $R$
in eq.~(\ref{rot2}), the constraints are~\cite{TO}
\bea
e^{\t(\Gamma_{12} + \Gamma_{34})} \Gamma_{091357}\e &=& \e, \nn
e^{-\t(\Gamma_{56} + \Gamma_{78})} \Gamma_{091357}\e &=& \e.
\label{cond3}
\ena
If we define
\EQ
\e' = e^{\shalf \t(\Gamma_{56} + \Gamma_{78})}\e,
\EN
where the independent numbers of components of $\e'$ and $\e$ are the same,
then the condition (\ref{cond3}) is rewritten as
\bea
\Gamma_{091357}\e' &=& \e', \nn
e^{\t(\Gamma_{12} + \Gamma_{34}+\Gamma_{56} + \Gamma_{78})}\e' &=& \e',
\ena
which is exactly the same as that in~\cite{TO}. Therefore, our T-dualized
configuration with restriction $B_{12}=B_{34}=-B_{56}=-B_{78}$ in
super Yang-Mills theory is the counterpart of the solution of $D=11$
supergravity with 3/16 supersymmetry found in~\cite{GG,TO}.

In the M(atrix) picture, our gauge configuration is the T-dual of the
configuration \{0222244\} with $n_{12}^{(l)} n_{34}^{(l)}/l,
n_{56}^{(k)} n_{78}^{(k)}/k$ 4-branes wrapped on (1234) and (5678)
directions; $n_{12}^{(l)}, n_{34}^{(l)}, n_{56}^{(k)}, n_{78}^{(k)}$
2-branes along (12), (34), (56) and (78) directions; and $N$ 0-branes
as always~\cite{TA2,GRT}.

Now we compare the energy of super Yang-Mills
theory with that expected of this configuration of branes in M(atrix)
theory. The $U(N)$ Hamiltonian gives
\EQ
H_{YM} = \frac{1}{4g_{YM9}^2}\; {\rm Tr} \int d^8 x F_{ij} F_{ij}
= \frac{\prod_{i=1}^8 a_i}{2 g_{YM9}^2} \left[l (B_{12}^2 + B_{34}^2)
 + k ( B_{56}^2 + B_{78}^2) \right].
\label{ham}
\EN

Noting that the configuration under discussion consists of two D4-branes
of mass $m_1$ and $m_2$ (defined as the two terms in eq.~(\ref{mass})),
the energy and 11-th momentum in M(atrix) theory in the infinite
momentum frame are given by
\bea
&& E = \frac{m_1^2}{2 p^{(1)}_{11}} + \frac{m_2^2}{2 p^{(2)}_{11}},\nn
&& P_{11} = p^{(1)}_{11} + p^{(2)}_{11} ; \;\;
p^{(1)}_{11} = \frac{l}{R_{11}} ; \;\;
p^{(2)}_{11} = \frac{k}{R_{11}} .
\ena

As discussed in~\cite{HT}, the field theory limit of the configuration
corresponds to taking $a_i$ to be of the same order of magnitude while
sending $\a'/a_2 a_4$ and $\a'/a_6 a_8$ to zero, and
$n_{12}^{(l)} n_{34}^{(l)}/l$ and $n_{56}^{(k)} n_{78}^{(k)}/k$
are of order 1~\cite{GR1}. Thus when we expand in $1/N$,
the contribution from $V_{2457}^{(l)}$ and $V_{1368}^{(k)}$ can be ignored.
Furthermore, in M(atrix) theory, the energy from 0-branes has not been
taken into account. In order to compare the energy of super Yang-Mills
theory with that in M(atrix) theory, the contribution from original 0-branes,
that is, $V_{1357}^{(l)}$ and $V_{1357}^{(k)}$ in the present case,
should be subtracted from $E$~\cite{GO,KK}. This procedure gives the energy
$E$ in the $N\to\infty$ limit as
\EQ
E=\frac{ l (B_{12}^2 + B_{34}^2) + k ( B_{56}^2 + B_{78}^2)}{4\pi g_s},
\EN
where we have taken $R_{11}=g_s \sqrt{\a'}$ and $g_8=g_s \prod_{i=1}^8 a_i$
with $g_s$ being the string coupling. With the identification of the
string coupling with the gauge coupling~\cite{GO,GRT}
\EQ
g_{YM9}^2 = 2 \pi g_s \prod_{i=1}^8 a_i,
\EN
we find
\EQ
E=\frac{\prod_{i=1}^8 a_i}{2 g_{YM9}^2} \left[l (B_{12}^2 + B_{34}^2)
 + k ( B_{56}^2 + B_{78}^2) \right],
\label{en}
\EN
in perfect agreement with the super Yang-Mills energy (\ref{ham}).

If $B_{12}=B_{34}=-B_{56}=-B_{78}$, the resulting configuration describes
BPS-saturated bound state which preserves 3/16 unbroken supersymmetry.
The match of energies in super Yang-Mills theory and in M(atrix)
theory indicates that the solution with 3/16 supersymmetry can be
realized in M(atrix) theory, which is T-dual of two sets of
longitudinal M5-branes (with common string direction) at angles.

So far, the explicit solution in super Yang-Mills theory, whose
T-dualized configuration corresponds to two sets of D4-branes with
relative angles, has been constructed. The gauge configuration
possesses 3/16 unbroken supersymmetry, and can be considered as the
counterpart of the solution of $D=11$ supergravity with the same amount
of supersymmetry as discussed in~\cite{GG,TO}. The energy of the
Born-Infeld theory has shown that the geometrical interpretation
of the T-dualized configuration as two sets of D4-branes at
angles is quite reasonable. Also it has been found that the
energy of super Yang-Mills theory coincides with that of
M(atrix) theory (in the infinite momentum frame with the
identification of the longitudinal momentum $P_{11}=N/R_{11}$).
These facts show that
the configuration of the bound state with 3/16 unbroken supersymmetry
can be realized in M(atrix) theory, which describes two sets of 
longitudinal M5-branes (with common string direction) at angles.

It is known that if the number of ND-directions is eight, a fundamental
string is created when two such orthogonal D4-branes cross~\cite{HW}.
With the above configuration at hand, one of interesting applications is
to study whether a similar fundamental string will be created or not
when two D4-branes at angles cross. Another speculation motivated by
the above discussion is that more
general solution in super Yang-Mills theory on $T^8$, where the
field strength is not necessarily constant and still describes
(by T-duality) two sets of D4-branes at angles, might be 
expected. The work along this line is probably related to~\cite{PAT}.
As the moduli space is important in determining the spectrum
of bound states, it would be interesting to discuss the moduli
space of the above configuration, which presumably corresponds to
eight-dimensional analogue of the well-studied moduli space of
self-dual connections on four-manifolds. These problems will be
further studied elsewhere.

\section*{Acknowledgement}
We would like to thank E. Keski-Vakkuri and P. Kraus for valuable
discussions on the evaluation of energy in the M(atrix) theory.
This work was supported in part by Grant-in-aid from the Ministry of
Education, Science, Sports and Culture No. 96208.

\newcommand{\NP}[1]{Nucl.\ Phys.\ {\bf #1}}
\newcommand{\AP}[1]{Ann.\ Phys.\ {\bf #1}}
\newcommand{\PL}[1]{Phys.\ Lett.\ {\bf #1}}
\newcommand{\NC}[1]{Nuovo Cimento {\bf #1}}
\newcommand{\CMP}[1]{Comm.\ Math.\ Phys.\ {\bf #1}}
\newcommand{\PR}[1]{Phys.\ Rev.\ {\bf #1}}
\newcommand{\PRL}[1]{Phys.\ Rev.\ Lett.\ {\bf #1}}
\newcommand{\PRE}[1]{Phys.\ Rep.\ {\bf #1}}
\newcommand{\PTP}[1]{Prog.\ Theor.\ Phys.\ {\bf #1}}
\newcommand{\PTPS}[1]{Prog.\ Theor.\ Phys.\ Suppl.\ {\bf #1}}
\newcommand{\MPL}[1]{Mod.\ Phys.\ Lett.\ {\bf #1}}
\newcommand{\IJMP}[1]{Int.\ Jour.\ Mod.\ Phys.\ {\bf #1}}
\newcommand{\JP}[1]{Jour.\ Phys.\ {\bf #1}}


\begin{thebibliography}{99}
\bibitem{POL} J. Polchinski, \PRL{75} (1995) 4724, hep-th/9510017;
 preprint, hep-th/9611050.
\bibitem{W1} E. Witten, \NP{B460} (1996) 335, hep-th/9510135.
\bibitem{BFS} T. Banks, W. Fischler, S.H. Shenker and L. Susskind,
 \PR{D55} (1997) 5112, hep-th/9610043.
\bibitem{GR1} Z. Guralnik and S. Ramgoolam, \NP{B499} (1997) 241,
 hep-th/9702099.
\bibitem{BD} M. Berkooz, M.R. Douglas and R.G. Leigh, \NP{B480} (1996) 265,
 hep-th/9606139.
\bibitem{BL} V. Balasubramanian and R.G. Leigh, \PR{D55} (1997) 6415,
 hep-th/9611165.
\bibitem{HT} A. Hashimoto and W. Taylor IV, preprint, hep-th/9703217.
\bibitem{GNS} E. Gava, K.S. Narain and M.H. Sarmadi, preprint, hep-th/9704006.
\bibitem{GO} R. Gopakumar, preprint, hep-th/9704030.
\bibitem{GR} Z. Guralnik and S. Ramgoolam, preprint, hep-th/9708089.
\bibitem{LI} G. Lifschytz, \NP{B499} (1997) 283, hep-th/9610125.
\bibitem{BM} J.C. Breckenridge, G. Michaud and R.C. Meyers, \PR{D55} (1997)
 6438, hep-th/9611174; preprint, hep-th/9703041;
 G. Michaud and R.C. Meyers, preprint, hep-th/9705079.
\bibitem{BC} K. Behrndt and M. Cveti\v{c}, \PR{D56} (1997) 1188,
 hep-th/9702205.
\bibitem{CC} M.S. Costa and M. Cveti\v{c}, preprint, hep-th/9703204.
\bibitem{HA} N. Hambli, preprint, hep-th/9703179.
\bibitem{GG} J.P. Gauntlett, G.W. Gibbons, G. Papadopoulos and P.K. Townsend,
 preprint, hep-th/9702202.
\bibitem{TO} P.K. Townsend, preprint, hep-th/9708074.
\bibitem{PT} G. Papadopoulos and P. Townsend, \PL{B380} (1996) 273,
 hep-th/9603087.
\bibitem{TS1} A.A. Tseytlin, \NP{B475} (1996) 149, hep-th/9604035;
 \NP{B487} (1997) 141, hep-th/9609212.
\bibitem{GKT} J.P. Gauntlett, D.A. Kastor and J. Traschen, \NP{B478} (1996)
 544, hep-th/9604179.
\bibitem{ZML} J.-G. Zhou, H.J.W. M\"{u}ller-Kirsten, J.-Q. Liang and
 F. Zimmerschild, \NP{B487} (1997) 155, hep-th/9611146.
\bibitem{AEH} R. Argurio, F. Englert and L. Houart, \PL{B398} (1997) 61,
 hep-th/9701042.
\bibitem{O} N. Ohta, \PL{B403} (1997) 218, hep-th/9702164.
\bibitem{TH} G. 't Hooft, \CMP{81} (1981) 267.
\bibitem{W2} E. Witten, \NP{B460} (1996) 541, hep-th/9511030.
\bibitem{DO} M. Douglas, preprint, hep-th/9512077.
\bibitem{TA1} W. Taylor IV, preprint, hep-th/9705116.
\bibitem{BKS} L. Baulieu, H. Kanno and I.M. Singer, preprint, hep-th/9704167.
\bibitem{HO} P. Horava, preprint, hep-th/9705055.
\bibitem{BSS} T. Banks, N. Seiberg and S. Shenker, \NP{B490} (1997) 91,
 hep-th/9612157.
\bibitem{TA2} W. Taylor IV, \PL{B394} (1997) 283, hep-th/9611042.
\bibitem{GRT} O.J. Ganor, S. Ramgoolam and W. Taylor IV, \NP{B492} (1997)
 191, hep-th/9611202.
\bibitem{KK} E. Keski-Vakkuri and P. Kraus, preprints, hep-th/9706196;
 hep-th/9709122.
\bibitem{HW} A. Hanany and E. Witten, preprint, hep-th/9611230;
 C.P. Bachas, M.R. Douglas and M.B. Green, preprint, hep-th/9705074;
 U. Danielsson, G. Ferreti and I.R. Klebanov, \PRL{79} (1997) 1984,
 hep-th/9705084;
 O. Bergman, M.R. Gaberdiel and G. Lifschytz, preprint, hep-th/9705130;
 S.P. de Alwis, preprint, hep-th/9706142;
 P.-M. Ho and Y.-S. Wu, preprint, hep-th/9708137.
\bibitem{PAT} G. Papadopoulos and A. Teschendorff, preprint, hep-th/9708116.
\end{thebibliography}
\end{document}